\documentstyle[12pt,epsfig]{article}

\begin{document}
\topmargin -1.4cm
\oddsidemargin -0.8cm
\evensidemargin -0.8cm 

\title{W condensation at the LHC?}

\vspace{1.5cm}

\author{P. Olesen\\
{\it The Niels Bohr International Academy}\\{\it  The Niels Bohr Institute}\\
{\it Blegdamsvej 17, Copenhagen \O, Denmark }}

\maketitle
\begin{abstract}
We discuss the possibility of the generation and subsequent decay 
of a $W$ and $Z$ condensate in LHC collisions. We point out that
a process like $h\rightarrow\gamma\gamma$, which involves a virtual
$W$ loop, may have an enhancement if the condensate exists. Even if this
is not the case, the $W$ propagator is influenced by the strong, but short
lived magnetic field generated in proton-proton collisions. This will influence
the di-gamma decay of the Higgs particle.
 
\end{abstract}


\thispagestyle{empty}

\vskip0.5cm

The major discovery of the Higgs particle \cite{exp1},\cite{exp2} gives 
ample opportunity for
analyzing the experimental data in order to see if there are clues for
deviations from the Standard Model (SM) \cite{p},\cite{f}. In particular, the 
preliminary data indicating that the two photon decay of the Higgs boson ($h$)
may deviate by a  factor 1.5-2 from the SM gives rize to new 
ideas \cite{p},\cite{f}. At the same time it must be understood
why the channels $h\rightarrow ZZ^*$ and $h\rightarrow WW^*$ seem to 
essentially agree with the SM. It may of course well be that the 
relatively modest excess in $h\rightarrow\gamma\gamma$
may diminish when all the data are analysed \cite{p}, so that in the
end the SM works perfectly.

In this note we presents the 
point of view that we should be sure that when the {\it perturbative} SM 
is used as a calibration of
the data, there are no modifications from non-perturbative 
condensates existing in the SM. For example, the process
$h\rightarrow \gamma\gamma$, which is calculated perturbatively \cite{1}-
\cite{8}, proceeds 
through loops involving {\it all} momenta for the
$W$ mesons (or quarks) which are, at least in principle, sensitive to 
background fields. If vector boson condensation occurs, it means that there 
are more vectors than dictated by perturbation theory, which is an
expansion around $W=Z=0$.  

In subsection 1 we discuss the arguments for a $W$ and $Z$ condensate, and
in subsection 2 we present arguments against such a condensate. In 3 we
present a brief discussion of the difference between $h\rightarrow\gamma\gamma$
and $h\rightarrow$vector-vector.

\vskip0.3cm

{\it 1. Arguments for a W and Z condensate. }

\vskip0.2cm

Within the SM there may be an ``anomalous'' behavior relative to 
perturbation theory
in the form of condensation of
$W$ and $Z$ mesons. This phenomenon was discussed a long time ago by 
Ambj\o rn and the  
author \cite{us} as a possibility in high energy collisions. The starting
point is that in the proton-proton collision large magnetic fields
are generated by quarks with charge $e_q$,
\begin{equation}
{\bf H}=\frac{e_q}{4\pi}~\frac{{\bf v\times r}}{r'^3\sqrt{1-v^2}}
\label{lhc}
\end{equation}
where
\begin{equation}
r'^2=b^2+r_L^2/(1-v^2).
\end{equation}
Here $v\sim 1$ and $\bf r$ is the vector from the simultaneous point of 
observation to the point charge. Also $b$ is the impact parameter and $r_L$
is the component of $\bf r$ in the direction of $\bf v$.

It has been known for a long time that magnetic fields exceeding the limit
\begin{equation}
H\geq \frac{m_W^2}{e}
\label{threshold}
\end{equation}
lead to an instability \cite{nkn} provided the field is homogeneous. 

The $H-$field (\ref{lhc}) easily exceeds the threshold (\ref{threshold}). 
For example in proton-proton scattering \cite{us}
\begin{eqnarray}
2H_{\rm crit}=\frac{m_w}{e}~~{\rm for}~~\sqrt{s}=420~{\rm GeV}~~
(e_q=e/3),\nonumber \\
=210~{\rm GeV}~~(e_q=2e/3).
\end{eqnarray}

Of course the field (\ref{lhc}) is not homogeneous, but it provides 
an environment with a very large background magnetic field, exceeding the 
threshold (\ref{threshold})  considerably. This is potentially
an unstable situation because the electroweak Lagrangian contains a term
\begin{equation}
-i(e f_{\mu\nu}^{\rm external}+g\cos\theta ~Z_{\mu\nu})~W_\mu^*W_\nu.
\label{term}
\end{equation}
When  the external magnetic field exceeds the threshold value (\ref{threshold})
this term  becomes tachyonic. For a $H$ field in the third direction the
eigenvalues are $\pm eH$ with eigenvector $(W_1,W_2)$. The negative eigenvalue 
exceeds the mass term $-m_w^2 W_\mu^*W_\mu$ and causes energy to be 
transferred from the $H$ field to the $W'$s (and $Z'$s). These fields
are expected to increase exponentially with time. An ultimate 
stabilization is expected from the $W^4$ term in the Lagrangian.For further 
discussion we refer to \cite{us}. 

The situation is more complicated than indicated by the simple minded 
considerations above. The magnetic field is time dependent, and the whole set 
up should
be investigated numerically. We mention that for the pure Yang-Mills
Lagrangian there has recently been numerical calculations \cite{montreal}
which shows that a large magnetic field in one color direction is unstable, and
energy is transferred to other degrees of freedom which grow exponentially in 
time as exp$(t\sqrt{gH})$.  This is valid for magnetic fields coherent over
a distance scale which is large compared to the Larmor scale $1/\sqrt{gH}$. 
It would of course be desirable if numerical calculations with a realistic
proton-proton field could be carried out in the electroweak case for the 
background field (\ref{lhc}).

It should be mentioned that if the background field is large enough then in the
static case ultimately the broken symmetry is restored \cite{restore},
\cite{r1}, \cite{restore2},
because the $W$ and $Z$ condensates become so large that the Higgs breaking
disappears. Whether this can happen in the time dependent case is not known,
but it is an interesting possibility, especially in view of the fact that the
field (\ref{lhc}) can be large enough for small impact parameters $b$ to
induce a transistion to the unbroken phase.

\vskip0.5cm

{\it 2. Arguments against a W and Z condensate}

\vskip0.2cm

Soon after the appearence of ref. \cite{us} a paper was published with the
conclusion that the proposed process of W-boson condensation in high-energy
p-p collisions does not work \cite{crit}. With the protons moving in the $z$ 
direction the
magnetic field was assumed to have an infinite extension in the $x-y-$plane, 
corresponding to afield
\begin{equation}
{\bf H}=\frac{e}{b}~\delta (r_L)~{{\bf e}_x},
\label{delta}
\end{equation}
where $b$ is the impact parameter. It is argued that the neglect of the 
fall-off of the field (\ref{lhc}) in the transverse direction is acceptable 
if the impact parameter is much larger than the size of the Landau orbits
in the magnetic field. The final
result of the calculations in \cite{crit} violates this condition, and
hence the negative conclusion follows.

In \cite{crit} the time dependence of the condensation has not been taken
into consideration, although this is obviously needed in
 a fully realistic treatment of the problem.
Another possible criticism of the critique in \cite{crit} is that the 
delta-function (\ref{delta})
is perhaps too structureless. The inverse impact parameter enter as a factor
in front of the delta function, but one also has an implicit $b$ dependence
in $\delta (0)\propto\gamma/b$. 

\vskip0.5cm

{\it 3. Consequences of a condensate: $h\rightarrow\gamma\gamma$}

\vskip0.2cm

 Bearing in mind that the existence of a $W$ 
and $Z$ condensate 
may only be marginally possible, we now end this note by some remarks on how
to see a condensate should it exist in spite of the complaints in \cite{crit}.

The possible existence of a  condensate means that $W'$s and $Z'$s are
extracted from the vacuum. The real vacuum then consists of non-trivial fields,
in contrast to the perturbative vacuum which is an expansion around $W=Z=0$.
Thus the condensate means that there should be vector fields which do not exist
perturbatively. This is true in the {\it infrared}, since the instability
occurs only when the relevant momenta are below $\sqrt{eH }> m_W$. 

The increase in the number of vector particles due to the condensate thus
only happens if infrared momenta are involved. 
The involvment of a range of (virtual) momenta happens in the process 
$h\rightarrow\gamma\gamma$ which proceeds through a 
loop \cite{1}-\cite{8}, thus integrating over {\it all}  loop momenta, 
including the infrared
crucial for condensation. The $W-$propagators involved in the loop
should be propagators calculated in the background of the strong
magnetic field. Thus, in the 
indirect decay of the Higgs particle to two photons the condensate
can have its full impact, provided of course that it exists. 

For the direct processes $h\rightarrow W^*W$ and $h\rightarrow Z^*Z$ the
situation is not so clear, since the momenta of the vectors decaying
from the Higgs particle are
totally fixed by the kinematics, in contrast to the di-photon decay where
the momentum is a loop variable. In the direct decay  there are no vector 
propagators, so the 
condensate will not
have a direct effect, even if it exists.

In conclusion we see that if there is a $W-$ and $Z-$condensate there
is a difference between the process $h\rightarrow\gamma\gamma$, where
the $W'$s can be extracted non-perturbatively through condensation,
and the direct vector decay, where no $W$ propagators are involved. The
important question is of course whether the condensate does exist. 
However, it should be emphasized that even if there is not enough time 
for condensation to occur, the {\it W-propagator is influenced by the
large magnetic field} (\ref{lhc}). The calculations of the di-gamma decay
\cite{1}-\cite{8} are based on the perturbative $W$ propagator, which is thus
not
entirely realistic, which may explain possible experimental deviations
in the decay $h\rightarrow\gamma\gamma$
from the calculations.

The 
most likely place where to find the effects discussed above
 is in heavy ion collisions.
In any case it is most likely that numerical studies are needed to settle the
question of the magnitude of the magnetic effect.


\begin{thebibliography}{X}
\bibitem{exp1}ATLAS Collaboration, http://cdsweb.cern.ch/record/1460439
\bibitem{exp2}CMS Collaboration,http://cdsweb.cern.ch/record/1460438?ln=en 
\bibitem{p}N. A. Hamed, K. Blum, R. T. D'Agnolo and J. J. Fan, arXiv:1297.4482
\bibitem{f}L. G. Almeida, E. Bertuzzo, P. A. N. Machado and R. Z. Funchal,
arXiv:1207.5254
\bibitem{1}L. Resnik, M. K. Sundaresan, P. J. S. Watson, Phys. Rev. D 8 (1973)
172
\bibitem{2}R. Ellis, M. K. Gaillard, and D. V. Nanopoulos, Nucl. Phys. B 106
(1976) 292
\bibitem{3}M. A. Shifman, A. I. Vainstein, M. B. Voloshin, and V. I. Zakharov, 
Sov. J. Nucl. Phys. 30 (1979) 711
\bibitem{4}G. Rizzo,  Phys. Rev. D22 (1980) 178
\bibitem{5}M. A. Shifman, A. I. Vainstein, M. B. Voloshin, and V. I. Zakharov,
Phys. Rev. D 85.045004; arXiv:1109.1785
\bibitem{6}R. Gastmans, S. L. Wu, and T. T. Wu, arXiv:1108.5322
\bibitem{7}R. Gastmans, S. L. Wu, and T. T. Wu, arXiv:1108.5872
\bibitem{8}D.  Huang, Y. Tang, and Y-L. Wu, Comm. Theor. Phys. 57 (2012)
427; arXiv:1109.4846 
\bibitem{us}J. Ambj\o rn and P. Olesen, Phys. Lett. B 257 (1991) 201.
\bibitem{nkn}N. K. Nielsen and P. Olesen, Nucl. Phys. B 144 (1978) 376
\bibitem{montreal}A. Kurkela and G. D. Moore, arXiv:1207.1663
\bibitem{restore} J. Ambj\o rn and P. Olesen, Nucl. Phys. B 315 (1989) 606
\bibitem{r1}J. Ambj\o rn and P. Olesen, Nucl. Phys. B 330 (1990) 193
\bibitem{restore2}M. N. Chemodub, J. Van Doorsselaere, and H. Verschelde, 
arXiv:1203.3071 
\bibitem{crit}S. Schramm, B. M\"uller, and A. J. Schramm, Phys. Lett. B 277
(1992) 512


\end{thebibliography}
\end{document}